\documentclass[10pt,twocolumn,conf]{IEEEtran}
\usepackage{times}
\usepackage{amsbsy}
\usepackage{latexsym}
\usepackage{amsmath}
\usepackage[ansinew]{inputenc}
\usepackage[dvips ]{graphicx}
\input epsf
\usepackage{epic,eepic}
\usepackage{url}
\usepackage{algorithm}
\usepackage{color}
\usepackage{booktabs}
\usepackage{setspace}
\usepackage{cite}
\IEEEoverridecommandlockouts

\title{\Huge Resource Allocation and Interference Mitigation Techniques for Cooperative Multi-Antenna and Spread Spectrum
Wireless Networks \vspace{0.05em}}
\author{Rodrigo C. de Lamare and Patrick J. Clarke  \\ Communications Research Group \\ Department of Electronics,
    University of York, York Y010 5DD, United Kingdom \\
    Emails: \protect\url{rcdl500@ohm.york.ac.uk}
\thanks{\footnotesize The work of the authors was
supported by the University of York, York Y010 5DD, United Kingdom.
}}

\begin{document}
\maketitle

\begin{abstract}

This chapter presents joint interference suppression and power
allocation algorithms for DS-CDMA and MIMO networks with multiple
hops and amplify-and-forward and decode-and-forward (DF) protocols.
A scheme for joint allocation of power levels across the relays
and linear interference suppression is proposed. We also consider
another strategy for joint interference suppression and relay
selection that maximizes the diversity available in the system.
Simulations show that the proposed cross-layer optimization
algorithms obtain significant
gains in capacity and performance over existing schemes.

\end{abstract}

\section{Introduction}

Multiple-antenna wireless communication systems can exploit the
spatial diversity in wireless channels, mitigating the effects of
fading and enhancing their performance and capacity. Due to the size
and cost of mobile terminals, it is considered impractical to equip
them with multiple antennas. However, spatial diversity gains can be
obtained when single-antenna terminals establish a distributed
antenna array via cooperation \cite{sendonaris}-\cite{laneman04}. {
The use of cooperative strategies can lead to several types of gains
\cite{dohler,pabst}, namely, pathloss, diversity and multiplexing
gains. Pathloss gains allow a significant reduction in the
transmitted power for an equivalent performance, can increase the
coverage \cite{ettefagh} and enhance the interference suppression
capability \cite{dohler,pabst}. The diversity gains improve the
performance of the wireless system with respect to the probability
of error because the transmission of multiple copies of the signals
reduce the probability that the message will not be received
correctly. The multiplexing gains \cite{so}, which correspond to the
additional number of bits that the system can transmit as compared
to a single-antenna link, can be obtained when a designer can use
relays to form independent channels and increase the rate of
communication.

Despite the many advantages in terms of gains as previously
outlined, cooperative communications also entail some disadvantages
such as signalling overheads \cite{pabst}, more computationally
complex scheduling algorithms \cite{dohler} and increased latency
\cite{lu}. For this reason, it is important to weigh the pros and
cons of cooperative techniques prior to their adoption and consider
the practical scenarios of interest \cite{kyritsi}. Motivated by
their performance and diversity gains, cooperative techniques are
now being considered for the next generation of mobile networks
\cite{Coop_comms_mobile_ad_hoc_networks_Scaglione,fan,peters}. In
cooperative systems, terminals or users relay signals to each other
in order to propagate redundant copies of the same signals to the
destination user or terminal. To this end, the designer must resort
to a cooperation protocol such as amplify-and-forward (AF)
\cite{laneman04}, decode-and-forward (DF) \cite{laneman04,huang} and
compress-and-forward (CF) \cite{kramer}.

In order to obtain the benefits of cooperative techniques,
designers must address a number of problems that are encountered
in cooperative wireless systems. These problems include physical-layer strategies
such as synchronization, interference mitigation, and parameter
estimation. However, designers also have to consider a number of associated
problems that belong to higher protocol layers and include the
allocation of resources such as power, relays and rate. These tasks
present an opportunity to perform cross-layer design and to obtain
very significant gains in performance and capacity for cooperative
wireless networks. This chapter is concerned with cross-layer
design techniques for cooperative wireless networks and investigates
the benefits of approaches that jointly mitigate interference
and perform resource allocation.

In this chapter, we will consider two types of schemes, namely,
direct-sequence code-division multiple access (DS-CDMA)
\cite{ziemer,honig}  and multi-input multi-output (MIMO)
\cite{foschini,telatar} systems. The former is of fundamental
importance in wireless ad-hoc and sensor networks \cite{dohler},
whereas the latter is one of the main ingredients of future wireless
cellular networks. When implementing cooperative techniques in
wireless systems, designers often consider the transmission
technologies available and their suitability to certain
applications. Therefore, the concept of distributed antenna arrays
can be easily extended to techniques such as MIMO
\cite{foschini,telatar} and DS-CDMA systems \cite{ziemer,honig}.

In the context of MIMO systems, one can obtain substantial
multiplexing \cite{foschini,telatar,vblast} and diversity gains
\cite{alamouti,tarokh} with the deployment of multiple antennas at
both ends of the wireless system. MIMO technology is poised to equip
most of the future wireless systems and can be incorporated in
conjunction with other transmission systems. There are two basic
configurations which exploit the nature of the wireless channel:
spatial multiplexing \cite{vblast} and diversity \cite{tarokh}.
Spatial multiplexing relies on the concept of forming individual
data stream between pais of transmit and receive antennas. The
capacity gains of spatial multiplexing grow linearly with the
minimum number of transmit and receive antennas
\cite{foschini,telatar} and allow a MIMO system to obtain a
considerable increase in data rates. Diversity configurations adopt
space-time codes \cite{alamouti,tarokh} to transmit data from the
antennas at the transmitter and can obtain a lower probability of
error.

DS-CDMA systems are a key multiple access technology for current and
future wireless communication systems. Such systems rely on the idea
of transmitting data with the aid of unique signatures, which are
also known as spreading codes. These signatures are responsible for
spreading the information in frequency, and allow the system to have
multiple users on the same channel. The advantages of DS-CDMA
include good performance in multi-path channels, flexibility in the
allocation of channels, increased capacity in bursty and fading
environments and the ability to share bandwidth with narrowband
communication systems without deterioration of either's systems
performance \cite{ziemer,honig}. Demodulating a desired user in a
DS-CDMA network requires processing the received signal in order to
mitigate different types of interference, namely, narrowband
interference (NBI), multi-access interference (MAI), inter-symbol
interference (ISI) and the noise at the receiver. The major source
of interference in most CDMA systems is MAI, which arises due to the
fact that users communicate through the same physical channel with
non-orthogonal signals.

The similarities between MIMO and CDMA systems include their
mathematically similar descriptions and their fundamental need for
interference mitigation. Indeed, the data streams of MIMO systems
operating in a spatial multiplexing configuration are equivalent to
the users of a DS-CDMA system. In order to separate data streams or
users, a designer must resort to detection techniques \cite{verdu},
which are very similar when applied to either MIMO or DS-CDMA. The
optimal maximum likelihood (ML) detector is often too complex to be
implemented for systems with a large number of antennas. For this
reason, designers often resort to suboptimal solutions that an
attractive trade-off between performance and complexity. These
include the sphere decoder (SD) algorithms \cite{hassibi}, linear
detectors \cite{verdu}, the successive interference cancellation
(SIC) approach \cite{vblast}, the parallel interference cancellation
(PIC) \cite{verdu} and the decision feedback (DF) detectors
\cite{delamaretc,choi} are techniques that can offer an attractive
trade-off between performance and complexity. These detection
algorithms can be combined with cross-layer design techniques for
enhanced interference mitigation and improved overall performance.
In this chapter, we are specifically interested in exploring the
advantages of linear detection with power allocation, data stream
and relay selection. }

\subsection{Prior and Related Work}

Prior work on cross-layer design for cooperative and multihop
communications has considered the problem of resource allocation
\cite{luo,long} in generic networks. These include power and rate
allocation strategies. Related work on cooperative multiuser DS-CDMA
networks has focused on the assessment of the impact of multiple
access interference (MAI) and intersymbol interference (ISI), the
problem of partner selection \cite{huang,venturino}, the bit error
ratio (BER) and outage performance analysis \cite{vardhe}, and
training-based joint power allocation and interference mitigation
strategies \cite{delamare_jpais,joung}. Previous works have also
considered the problem of antenna selection, relay selection (RS)
and diversity maximization, which are central themes in the MIMO
relaying literature \cite{Greedy_ant_selection_MIMO_relay_Ding,
joint_source_relay_opt_MIMO_relay_Koshy,
MIMO_antenna_discrete_optimization_Krishnamurthy}. However, current
approaches are often limited to stationary, single relay systems and
channels which assume the direct path from the source to the
destination is negligible
\cite{joint_source_relay_opt_MIMO_relay_Koshy}.

Most of these resource allocation and interference mitigation
strategies require a higher computational cost to implement the
power allocation and a significant amount of signalling, decreasing
the spectral efficiency of cooperative networks. This problem is
central to ad-hoc and sensor networks \cite{souryal} that employ
spread spectrum systems and require multiple hops to communicate
with nodes that are far from the source node. This is also of
paramount importance in cooperative cellular networks.

\subsection{Contributions}

In this chapter, we present joint interference suppression and power
allocation algorithms for DS-CDMA and MIMO networks with multiple
hops and AF and DF protocols. A scheme that jointly considers the
power allocation across the relays subject to group-based power
constraints and the design of linear receivers for interference
suppression is proposed. The idea of a group-based power allocation
constraint is shown to yield close to optimal performance, while
keeping the signalling and complexity requirements low. A
constrained minimum mean-squared error (MMSE) design for the receive
filters and the power allocation vectors is developed along with an
MMSE channel estimator for the cooperative system under
consideration. The linear MMSE receiver design is adopted due to its
mathematical tractability and good performance. { However, the
incorporation of more sophisticated detection strategies including
interference cancellation with iterative decoding \cite{delamaretc}
and advanced parameter estimation methods \cite{jidf} is also
possible}. In order to solve the proposed optimization problem
efficiently, a method to form an effective group of users and an
alternating optimization strategy are presented with recursive
alternating least squares (RALS) algorithms for estimating the
parameters of the receiver, the power allocation and the channels. A
joint relay selection and transmit diversity selection strategy for
MIMO networks with linear receivers is also proposed which optimizes
relay transmissions with minimal feedback requirements. Effectively
a novel approach to 1-bit power allocation, two joint discrete
optimization functions are formed which are solved using discrete
stochastic algorithms.

\subsection{Organisation of the Chapter}

The chapter is organized as follows. Section II describes
cooperative DS-CDMA and MIMO system models with multiple hops. Section III
formulates the problem, details the constrained MMSE design of the
receive filters and the power allocation vectors subject to a
group-based power allocation constraint, and describes an MMSE
channel estimator. An extension to cooperative MIMO systems
is also presented and discrete optimization problems are
formulated to jointly select the optimal relays and their
transmit antennas. Section IV presents an algorithm to form the
group and the alternating optimization strategy along with
RLS-type algorithms for estimating the parameters of the receiver,
the power allocation and the channels. For the solution of
the combinatorial problems posed by the relay selection
strategy, a pair of discrete stochastic algorithms are
introduced and their joint operation detailed. Section
V presents and discusses the simulation results and
Section VI draws the conclusions of this work.

\section{System and Data Models of Cooperative Wireless Systems}

{  In this section, we consider system and data models of
cooperative wireless systems. The basic idea is to use a linear
algebra approach to describe models of the cooperative systems of
interest. In particular, we focus on DS-CDMA and MIMO systems and we
present a unified approach to the description of these systems.}

\subsection{Cooperative DS-CDMA System and Data Model}

\begin{figure}[!htb]
\begin{center}
\def\epsfsize#1#2{1.0\columnwidth}
\hspace{-1em}\epsfbox{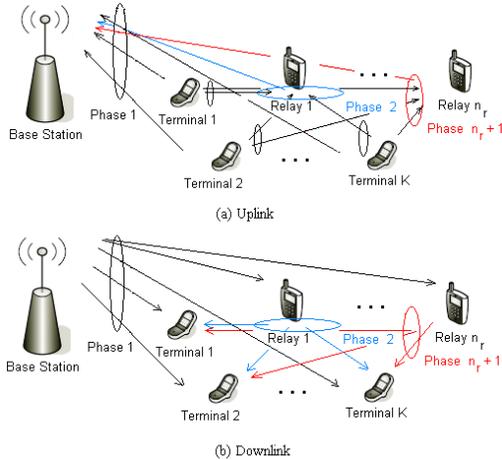} \vspace{-1.5em}\caption{(a) Uplink
and (b) downlink of the cooperative DS-CDMA system.} \label{figsys}
\end{center}
\end{figure}

Let us first consider a synchronous DS-CDMA network with multipath
channels. The DS-CDMA system operates with QPSK modulation, $K$
users, $N$ chips per symbol and $L$ as the maximum number of
propagation paths for each link. An outline of the system is
depicted in (\ref{figsys}). The system is equipped with AF and DF
protocols that allow communication in multiple hops using $n_r$
fixed relays in a repetitive fashion. We assume that the source node
or terminal transmits data organized in packets with $P$ symbols,
there is enough control data to coordinate transmissions and
cooperation, and the linear receivers at the relay and destination
terminals are synchronized with their desired signals. The received
signals are filtered by a matched filter, sampled at chip rate and
organized into $M \times 1$ vectors ${\boldsymbol r}_{sd}$,
${\boldsymbol r}_{sr_j}$ and ${\boldsymbol r}_{r_jd}$, which
describe the signal received from the source to the destination, the
source to the relays, and the relays to the destination,
respectively,
\begin{equation}
\begin{split}
{\boldsymbol r}_{sd} & = \sum_{k=1}^K  a_{sd}^k {\boldsymbol C}_k
{\boldsymbol h}_{sd,k}b_k   + {\boldsymbol \eta}_{sd}
\\ & \quad  + {\boldsymbol n}_{sd},
\\
{\boldsymbol r}_{sr_j} & = \sum_{k=1}^K a_{sr_j}^k {\boldsymbol C}_k
{\boldsymbol h}_{sr_j,k} {b}_k  + {\boldsymbol \eta}_{sr_j}
\\ & \quad + {\boldsymbol n}_{sr_1j},
\\
{\boldsymbol r}_{r_jd} & = \sum_{k=1}^K a_{r_jd}^k {\boldsymbol C}_k
{\boldsymbol h}_{r_jd,k} \tilde{b}_k  + {\boldsymbol \eta}_{r_jd}
\\ &  \quad+ {\boldsymbol n}_{r_jd},
\\ j & = 1, \ldots,n_r,~  i =1, \ldots, P \label{rvec}
\end{split}
\end{equation}
where $M=N+L-1$, $P$ is the number of packet symbols, $n_p=n_r+1$ is
the number of transmission phases or hops, and $n_r$ is the number
of relays. The vectors ${\boldsymbol n}_{sd}$, ${\boldsymbol
n}_{sr_j}$ and ${\boldsymbol n}_{r_jd}$ are zero mean complex
Gaussian vectors with variance $\sigma^2$ generated at the receivers
of the destination and the relays from different links, and the
vectors ${\boldsymbol \eta}_{sd}$, ${\boldsymbol \eta}_{sr_j}$ and
${\boldsymbol \eta}_{r_jd}$ represent the intersymbol interference
(ISI). The amplitudes of the source to destination, source to relay
and relay to destination  links for user $k$ are denoted by
$a_{sd}^k$, $a_{sr_j}^k$ and $a_{r_jd}^k$, respectively. The
quantities ${b}_k$ and $\tilde{b}_k$ represent the original and
reconstructed symbols by the AF or DF protocol at the relays,
respectively. The $M \times L$ matrix ${\boldsymbol C}_k$ contains
versions of the signature sequences of each user shifted down by one
position at each column as described by
\begin{equation}
{\boldsymbol C}_k = \left[\begin{array}{c c c }
c_{k}(1) &  & {\bf 0} \\
\vdots & \ddots & c_{k}(1)  \\
c_{k}(N) &  & \vdots \\
{\bf 0} & \ddots & c_{k}(N)  \\
 \end{array}\right],
\end{equation}
where ${\boldsymbol c}_k = \big[c_{k}(1), ~c_{k}(2),~ \ldots,~
c_{k}(N) \big]$ stands for the signature sequence of user $k$, the
$L \times 1$ channel vectors  from source to destination, source to
relay, and relay to destination are ${\boldsymbol h}_{sd,k}$,
${\boldsymbol h}_{sr_j,k}$, ${\boldsymbol h}_{r_jd,k}$,
respectively. By collecting the data vectors in (\ref{rvec})
(including the links from relays to the destination) into a $J
\times 1$ received vector at the destination, where $J=(n_r+1)M$, we
obtain {
\begin{equation}
\begin{split}
 \hspace{-0.5em}\left[\begin{array}{c} \hspace{-0.5em}
  {\boldsymbol r}_{sd} \\
 \hspace{-0.5em} {\boldsymbol r}_{r_{1}d} \\
\hspace{-0.5em}  \vdots \\
 \hspace{-0.5em} {\boldsymbol r}_{r_{n_r}d}
\end{array}\right] & = \left[\begin{array}{c}
  \sum_{k=1}^K  a_{sd}^k {\boldsymbol C}_k {\boldsymbol h}_{sd,k}b_k \\
  \sum_{k=1}^K  a_{{r_1}d}^k {\boldsymbol C}_k {\boldsymbol h}_{{r_1}d,k}{\tilde b}_k^{{r_1}d} \\
  \vdots \\
  \sum_{k=1}^K  a_{{r_{n_r}}d}^k {\boldsymbol C}_k {\boldsymbol h}_{r_{n_r}d,k}{\tilde b}_k^{{r_{n_r}}d}
\end{array} \hspace{-0.5em} \right] \\ & \quad + {\boldsymbol \eta} + {\boldsymbol n}
\end{split}
\end{equation}}
Rewriting the above signals in a compact form yields
\begin{equation}
\begin{split}
{\boldsymbol r}[i] & = \sum_{k=1}^{K}  \widetilde{\boldsymbol
B}_k[i] \widetilde{\boldsymbol A}_k[i]
\underbrace{\widetilde{\boldsymbol {\mathcal C}}_k {\boldsymbol
h}_k[i]}_{{\boldsymbol p}_k[i]}+ {\boldsymbol \eta}[i] +
{\boldsymbol n}[i]
\\ & = \sum_{k=1}^{K}   \widetilde{\boldsymbol
B}_k[i] \widetilde{\boldsymbol A}_k[i] \widetilde{\boldsymbol
{\mathcal C}}_k {\boldsymbol  h}_k[i]+ {\boldsymbol \eta}[i] +
{\boldsymbol n}[i]
\\ & = \sum_{k=1}^{K}  {\boldsymbol P}_k[i]
{\boldsymbol B}_k[i] {\boldsymbol a}_k[i]+ {\boldsymbol \eta}[i] +
{\boldsymbol n}[i] 
, \label{recdata}
\end{split}
\end{equation}
where the $J \times (n_r+1)L$ matrix
$\widetilde{\boldsymbol {\mathcal C}}_k = {\rm diag} \{
{\boldsymbol C}_k \ldots {\boldsymbol C}_k \}$ contains copies of
${\boldsymbol C}_k$ shifted down by $M$ positions for each group of $L$ columns and zeros elsewhere. 
The $Q \times 1$ vector ${\boldsymbol h}_k[i]$, where $Q=(n_r+1)L$
contains the channel gains of the links between the source, the
relays and the destination, and ${\boldsymbol p}_k[i]$ is the
effective signature for user $k$. The $(n_r+1) \times (n_r+1)$
diagonal matrix ${\boldsymbol B}_k[i] = {\rm diag}(b_k[i]~ {\tilde
b}_k^{{r_1}d}[i] \ldots {\tilde b}_k^{{r_n}d}[i]) $ contains the
symbols transmitted from the source to the destination ($b_k[i]$)
and the $n_r$ symbols transmitted from the relays to the destination
(${\tilde b}_k^{{r_1}d}[i] \ldots {\tilde b}_k^{{r_n}d}[i]$) on the
main diagonal, and the $J \times J$ diagonal matrix
$\widetilde{\boldsymbol B}_k[i] = {\rm diag}(b_k[i]\bigotimes
{\boldsymbol I}_M~ {\tilde b}_k^{{r_1}d}[i]\bigotimes {\boldsymbol
I}_M \ldots {\tilde b}_k^{{r_n}d}[i]\bigotimes {\boldsymbol I}_M)$,
where $\bigotimes$ denotes the Kronecker product and ${\boldsymbol
I}_M$ is an identity matrix with dimension $M$. The $(n_r+1) \times
1$ power allocation vector ${\boldsymbol
a}_k[i]=[a_{sd}^k~a_{{r_1}d}^k\ldots a_{{r_{n_r}}d}^k]^T$ has the
amplitudes of the links, the $(n_r+1) \times (n_r+1)$ diagonal
matrix ${\boldsymbol A}_k[i]$ is given by ${\boldsymbol A}_k[i] =
{\rm diag} \{ {\boldsymbol a}_k[i] \}$, and the $J \times J$
diagonal matrix $\widetilde{\boldsymbol A}_k[i]= [a_{sd}^k\bigotimes
{\boldsymbol I}_M~a_{{r_1}d}^k\bigotimes {\boldsymbol I}_M \ldots
a_{{r_{n_r}}d}^k\bigotimes {\boldsymbol I}_M]^T $. The $J \times
(n_r+1)$ matrix ${\boldsymbol P}_k$ has copies of the effective
signature ${\boldsymbol p}_k[i]$ shifted down by $M$ positions for
each column and zeros elsewhere. The $J \times 1$ vector
${\boldsymbol \eta}[i]$ represents the ISI terms and the $J \times
1$ vector ${\boldsymbol n}[i]$ has the noise components.

\subsection{Cooperative MIMO System and Data Model}

\begin{figure}[!hb]
\begin{center}
\def\epsfsize#1#2{1.0\columnwidth}
\epsfbox{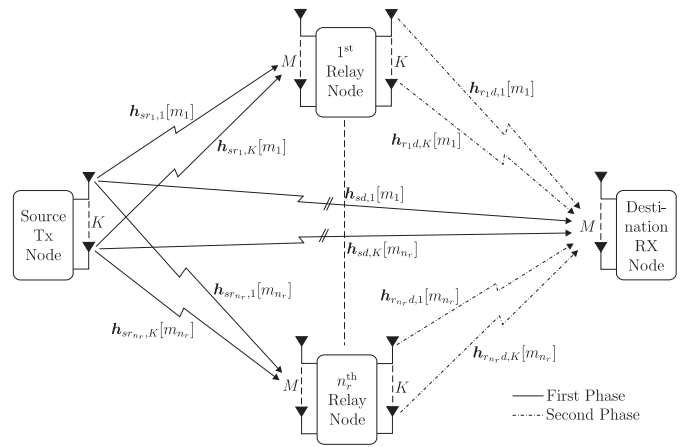}
\caption{Block diagram of $2$-phase cooperative MIMO system.} \label{figsys_mimo}
\end{center}
\end{figure}

{  Let us now consider a synchronous MIMO system model, which has
similarities with the DS-CDMA system model of the previous
subsection. We consider a narrowband MIMO system with flat fading
channels, QPSK modulation, $K$ transmit antennas, and $M$ receive
antennas as illustrated in Fig. \ref{figsys_mimo}.} The cooperative
MIMO network is equipped with DF protocol that allows communication
in $n_p=2$ hops using $n_r$ fixed relays in a repetitive fashion
where a non-negligible, direct source to destination link exists
during the first phase. We assume that the source node or terminal
transmits data organized in packets with $P$ symbols, there is
enough control data to coordinate transmissions and cooperation, and
the linear receivers at the relay and destination terminals are
synchronized with their desired signals. It should be noted that the
MIMO and CDMA system and data models are mathematically equivalent
and the main difference is that we employ for the MIMO version a
spreading code matrix ${\boldsymbol C}_k=1$.

The received signals are filtered by a matched filter, sampled at
chip rate and organized into $M \times 1$ vectors ${\boldsymbol
r}_{sd}$, ${\boldsymbol r}_{sr_j}$ and ${\boldsymbol r}_{rd}$, which
describe the signal received from the source to the destination, the
source to the relays, and the relays to the destination,
respectively,
\begin{equation}
\begin{split}
{\boldsymbol r}_{sd} & = \sum_{k=1}^K  a_{sd}^k
 {\boldsymbol h}_{sd,k}b_k[i]   +
{\boldsymbol \eta}_{sd} \\ & \quad  + {\boldsymbol n}_{sd},
\\
{\boldsymbol r}_{sr_j} & = \sum_{k=1}^K a_{sr_j}^k {\boldsymbol
h}_{sr_j,k} {b}_k[i]  + {\boldsymbol \eta}_{sr_j} \\ & \quad +
{\boldsymbol n}_{sr_j},
\\
{\boldsymbol r}_{r_jd} & = \sum_{k=1}^K a_{r_jd}^k
 {\boldsymbol h}_{r_jd,k} \tilde{b}_k[i]  +
{\boldsymbol \eta}_{r_jd} \\ &  \quad+ {\boldsymbol n}_{r_jd},
\\ j & = 1, \ldots,n_r,~  i =1, \ldots, P, ~ p= 1,2 \label{rvec}
\end{split}
\end{equation}
where $P$ is the number of packet symbols, $n_p=2$ is the number of
transmission phases or hops, and $n_r$ is the number of relays. The
vectors ${\boldsymbol n}_{sd}$, ${\boldsymbol n}_{sr_j}$ and
${\boldsymbol n}_{r_jd}$ are zero mean complex Gaussian vectors with
variance $\sigma^2$ generated at the receivers of the destination
and the relays from different links, and the vectors ${\boldsymbol
\eta}_{sd}$, ${\boldsymbol \eta}_{sr_j}$ and ${\boldsymbol
\eta}_{r_jd}$ represent the intersymbol interference (ISI).

The amplitudes of the source to destination, source to relay and
relay to destination  links for user $k$ are denoted by $a_{sd}^k$,
$a_{sr_j}^k$ and $a_{r_jd}^k$, respectively. The quantities
${b}_k[i]$ and $\tilde{b}_k[i]$ represent the original and
reconstructed symbols by the AF or DF protocol at the relays,
respectively. The $M \times 1$ spatial channel vectors from source
to destination, source to relay, and relay to destination are
${\boldsymbol h}_{sd,k}$, ${\boldsymbol h}_{sr_j,k}$, ${\boldsymbol
h}_{r_jd,k}$, respectively. By collecting the data vectors in
(\ref{rvec}) (including the links from relays to the destination)
into a $J \times 1$ received vector at the destination, where $J=n_p
M$ for MIMO systems, we obtain {
\begin{equation}
\begin{split}
 {\boldsymbol r}[i] & = \left[\begin{array}{c}
  \sum_{k=1}^K  a_{sd}^k  {\boldsymbol h}_{sd,k}b_k[i] \\
   \sum_{j}^{n_r} \sum_{k=1}^K  a_{{r_{n_r}}d}^k {\boldsymbol h}_{r_{n_r}d,k}{\tilde b}_k^{{r_{n_r}}d}[i]
\end{array} \hspace{-0.5em} \right] \\ & \quad  + {\boldsymbol n}[i]
\end{split}
\end{equation}}
Rewriting the above signals in a compact form yields
\begin{equation}
\begin{split}
{\boldsymbol r}[i] &  = \sum_{k=1}^{K n_r}
\widetilde{\boldsymbol B}_k[i] \widetilde{\boldsymbol A}_k[i]
{\boldsymbol  h}_k[i]+ {\boldsymbol n}[i]
\\ & = \sum_{k=1}^{K}  {\boldsymbol P}_k[i]
{\boldsymbol B}_k[i] {\boldsymbol a}_k[i]+
{\boldsymbol n}[i] 
, \label{recdata}
\end{split}
\end{equation}
The $J \times 1$ vector ${\boldsymbol h}_k[i]$ contains the spatial
channel gains of the links between the source, the relays and the
destination. The $n_p \times n_p$ diagonal matrix ${\boldsymbol
B}_k[i] = {\rm diag}(b_k[i]~ {\tilde b}_k^{{r_k}d}[i] ) $, for $1
\leq k \leq K$, and ${\boldsymbol B}_k[i] = {\rm diag}(0~ {\tilde
b}_k^{{r_k}d}[i] ) $, for $K < k \leq K n_r$, contains the symbols
transmitted from the source to the destination ($b_k[i]$) and the
$n_r$ symbols transmitted from the relays to the destination
(${\tilde b}_k^{{r_k}d}[i]$) on the main diagonal, and the $J \times
J$ diagonal matrix $\widetilde{\boldsymbol B}_k[i] = {\rm
diag}(b_k[i]\bigotimes {\boldsymbol I}_M~ {\tilde
b}_k^{{r_k}d}[i]\bigotimes {\boldsymbol I}_M )$, for $1 \leq k \leq
K$, and $\widetilde{\boldsymbol B}_k[i] = {\rm diag}(0 \bigotimes
{\boldsymbol I}_M~ {\tilde b}_k^{{r_k}d}[i]\bigotimes {\boldsymbol
I}_M ) $, for $K < k \leq K n_r$,. The $ n_p \times 1$ power
allocation vector ${\boldsymbol a}_k[i]=[a_{sd}^k~\ldots~ a_{r_{k}
d}^k]^T$ has the amplitudes of the links, the $n_p \times n_p $
diagonal matrix ${\boldsymbol A}_k[i]$ is given by ${\boldsymbol
A}_k[i] = {\rm diag} \{ {\boldsymbol a}_k[i] \}$, and the $J \times
J$ diagonal matrix $\widetilde{\boldsymbol A}_k[i]=
[a_{sd}^k\bigotimes {\boldsymbol I}_M~ \ldots~ a_{{r_{n_k}}d}^k
\bigotimes {\boldsymbol I}_M ]^T $. The $J \times n_p$ matrix
${\boldsymbol P}_k$ has copies of the spatial signatures
${\boldsymbol h}_k[i]$ shifted down by $M$ positions for each column
and zeros elsewhere. The $J \times 1$ vector ${\boldsymbol n}[i]$
has the noise components.

\section{Joint MMSE Receiver Design, Power Allocation, Relay Selection and Channel Estimation}

{  In this section, our aim is to describe techniques to mitigate
interference and allocate the power and select the best relays
according to the mean square error (MSE) criterion. We present a
joint receiver design and power allocation strategy using
constrained linear MMSE estimation and group-based power constraints
along with a linear MMSE channel estimator. The interesting aspect
of the group-based power constraints is that a designer can choose a
subset of users or data streams for power adjustment. Another
technique that is detailed here is a method called transmit
diversity selection (TDS) which operates with the linear MMSE
receivers. With the TDS technique the relay selection is used to
jointly optimize the selection of antennas in a strategy equivalent
to a $1$-bit transmit antenna power allocation.}

{  In order to describe the techniques necessary for interference
mitigation and resource allocation, we introduce an alternative way
of expressing the $J \times 1$ received vector in (\ref{recdata}).
Our goal is to separate the subset of users or data streams that
will be used for resource allocation with the group-based power
constraints. The modified $J \times 1$ received vector can be
expressed as }
\begin{equation}
{\boldsymbol r}[i] = {\boldsymbol P}_{\mathbf {\mathcal S}}[i]
{\boldsymbol B}_{\mathbf {\mathcal S}}[i] {\boldsymbol
a}_{{\mathbf {\mathcal S}},k}[i] + \sum_{k \neq {\mathbf {\mathcal
S}}} {\boldsymbol P}_k[i] {\boldsymbol B}_k[i] {\boldsymbol
a}_k[i]+ {\boldsymbol \eta}[i] + {\boldsymbol n}[i],
\label{recdatag}
\end{equation}
where ${\mathbf {\mathcal S}} = \{{\mathcal S}_1, {\mathcal S}_2,
\ldots, {\mathcal S}_G \}$ denotes the group of $G$ users to
consider in the design. The $J \times G(n_r+1)$ matrix ${\boldsymbol
P}_{\mathbf {\mathcal S}} = [ {\boldsymbol P}_{{\mathcal S}_1} ~
{\boldsymbol P}_{{\mathcal S}_2} ~ \ldots ~ {\boldsymbol
P}_{{\mathcal S}_G} ]$ contains the $G$ effective signatures of the
group of users. The $G(n_r+1) \times G(n_r+1)$ diagonal matrix
${\boldsymbol B}_{\mathbf {\mathcal S}}[i] = {\rm diag}(b_{{\mathcal
S}_1}[i]~ {\tilde b}_{{\mathcal S}_1}^{{r_1}d}[i] \ldots {\tilde
b}_{{\mathcal S}_1}^{{r_n}d}[i]~ \ldots ~ b_{{\mathcal S}_G}[i]~
{\tilde b}_{{\mathcal S}_G}^{{r_1}d}[i] \ldots {\tilde b}_{{\mathcal
S}_G}^{{r_n}d}[i])$ contains the symbols transmitted from the
sources to the destination and from the relays to the destination of
the $G$ users in the group on the main diagonal, the $G(n_r+1)
\times 1$ power allocation vector ${\boldsymbol a}_{{\mathbf
{\mathcal S}},k}[i]=[a_{sd}^{{\mathcal
S}_1}[i]~a_{{r_1}d}^{{\mathcal S}_1}[i] \ldots
a_{{r_{n_r}}d}^{{\mathcal S}_1}[i],~ \ldots, ~ a_{sd}^{{\mathcal
S}_G}[i]~a_{{r_1}d}^{{\mathcal S}_G}[i] \ldots
a_{{r_{n_r}}d}^{{\mathcal S}_G}[i]]^T$ of the amplitudes of the
links used by the $G$ users or data streams in the group.

\subsection{Linear MMSE Receiver Design and Power Allocation Scheme with
Group-Based Constraints}

The linear MMSE interference mitigation for user or data stream $k$
is performed by the receive filter ${\boldsymbol w}_k[i]=[
{w}_{k,1}[i],~ \ldots, ~ {w}_{k,J}[i]]$ with $J$ coefficients on the
received data vector ${\boldsymbol r}[i]$ and yields
\begin{equation}
z_k[i] = {\boldsymbol w}_k^H[i] {\boldsymbol r}[i],
\end{equation}
where $z_k[i]$ is an estimate of the symbols, which are processed
by a slicer $Q(\cdot)$ that performs detection and obtains the
desired symbol as $\hat{b}_k[i] = Q (z_k[i])$.

{  Let us now detail the linear MMSE-based design of the receivers
for user or data stream $k$ represented by ${\boldsymbol w}_k[i]$
and for the computation of the $G(n_r +1) \times 1$ power allocation
vector ${\boldsymbol a}_{{\mathbf {\mathcal S}},k}[i]$. This problem
can be cast as the following constrained optimization}
\begin{equation}
\begin{split}
[ {\boldsymbol w}_{k}^{\rm opt}, ~{\boldsymbol a}_{{{\mathbf
{\mathcal S}},k}}^{\rm opt}  ] & = \arg \min_{{\boldsymbol
w}_k[i], {\boldsymbol a}_{{\mathbf {\mathcal S}},k}[i]} ~
E[ (|{ b}_k[i] - {\boldsymbol w}^H_k[i]{\boldsymbol r}[i] |^2  ] \\
& {\rm subject ~to~}  {\boldsymbol a}_{{\mathbf {\mathcal
S}},k}^H[i] {\boldsymbol a}_{{\mathbf {\mathcal S}},k}[i] = P_{G},
\label{probg}
\end{split}
\end{equation}
{  In order to obtain expressions for the receive filter
${\boldsymbol w}_{k}[i]$ and the power allocation vector
${\boldsymbol a}_{{\mathbf {\mathcal S}},k}[i]$ subject to the
group-based power constraints, we need the help of the method of
Lagrange multipliers (\ref{probg}) \cite{haykin} that transforms a
constrained optimization into an unconstrained one. The MMSE
expressions for ${\boldsymbol w}_{k}[i]$ and ${\boldsymbol
a}_{{\mathbf {\mathcal S}},k}[i]$ are given by} {\small
\begin{equation}
\begin{split}
{\mathcal L}_k & = E\big[ |b_k[i] - {\boldsymbol w}_k^H[i] \big(
{\boldsymbol P}_{\mathbf {\mathcal S}}[i] {\boldsymbol B}_{\mathbf
{\mathcal S}}[i] {\boldsymbol a}_{{\mathbf {\mathcal S}},k}[i] \\
& \quad  + \sum_{k \neq {\mathbf {\mathcal S}}} {\boldsymbol
P}_k[i] {\boldsymbol B}_k[i] {\boldsymbol a}_k[i]   + {\boldsymbol
\eta}[i] + {\boldsymbol n}[i]\big) |^2 \big]   \\
& \quad + \lambda_k ({\boldsymbol a}_{{\mathbf {\mathcal S}},k}[i]
- P_{G}) , \label{lagt}
\end{split}
\end{equation}}
where $\lambda_k$ is a Lagrange multiplier.

{  Since the Lagrangian in (\ref{lagt}) is a function of both
${\boldsymbol w}_{k}[i]$ and ${\boldsymbol a}_{{\mathbf {\mathcal
S}},k}[i]$, we need to employ a strategy for optimization the
function with respect to both parameter vectors. The main idea is to
fix one of the parameter vectors and compute the gradient terms with
respect to the other parameter vector that minimizes the Lagrangian
and obtain the expression of interest. In particular, an expression
for ${\boldsymbol a}_{{\mathbf {\mathcal S}},k}[i]$ is obtained by
fixing ${\boldsymbol w}_k[i]$, taking the gradient terms of the
Lagrangian and equating them to zero, which yields}
\begin{equation}
{\boldsymbol a}_{{\mathbf {\mathcal S}},k}[i] = ( {\boldsymbol
R}_{{\mathbf {\mathcal S}},k}[i] + \lambda_k {\boldsymbol I})^{-1}
{\boldsymbol p}_{{\mathbf {\mathcal S}},k}[i] \label{avect}
\end{equation}
where the $G(n_r+1) \times G(n_r+1)$ covariance matrix
${\boldsymbol R}_{{\mathbf {\mathcal S}},k}[i] = E[ {\boldsymbol
B}_{\mathbf {\mathcal S}}^H[i]{\boldsymbol P}_{\mathbf {\mathcal
S}}^H[i]  {\boldsymbol w}_k[i] {\boldsymbol w}^H_k[i]{\boldsymbol
P}_{\mathbf {\mathcal S}}[i] {\boldsymbol B}_{\mathbf {\mathcal
S}}[i]]$  and the vector ${\boldsymbol p}_{{\mathbf {\mathcal
S}},k}[i] = E[b_k[i] {\boldsymbol B}_{\mathbf {\mathcal
S}}^H[i]{\boldsymbol P}_{\mathbf {\mathcal S}}^H[i]  {\boldsymbol
w}_k[i]]$ is a $G(n_r+1) \times 1$ cross-correlation vector. The
Lagrange multiplier $\lambda_k$ plays the role of a regularization
term and has to be determined numerically due to the difficulty of
evaluating its expression.

{  In order to compute the expression for ${\boldsymbol w}_k[i]$, we
fix ${\boldsymbol a}_{{\mathbf {\mathcal S}},k}[i]$, calculate the
gradient terms of the Lagrangian and equate them to zero which leads
to}
\begin{equation}
{\boldsymbol w}_k[i] = {\boldsymbol R}^{-1}[i] {\boldsymbol
p}_k[i], \label{wvect}
\end{equation}
where the covariance matrix of the received vector is given by
${\boldsymbol R}[i] = E[{\boldsymbol r}[i]{\boldsymbol r}^H[i]]$
and ${\boldsymbol p}_k[i] = E[b_k^*[i] {\boldsymbol r}[i]] $ is a
$J \times 1$ cross-correlation vector. The quantities
${\boldsymbol R}[i]$ and ${\boldsymbol p}_k[i]$ depend on the
power allocation vector ${\boldsymbol a}_{{\mathbf {\mathcal
S}},k}[i]$. The expressions in (\ref{avect}) and (\ref{wvect}) do
not have a closed-form solution as they have a dependence on each
other. Moreover, the expressions also require the estimation of
the channel vector ${\boldsymbol h}_k[i]$. Thus, it is necessary
to iterate (\ref{avect}) and (\ref{wvect}) with initial values to
obtain a solution and to estimate the channel. The network has to
convey the information from the group of users which is necessary
to compute the group-based power allocation including the filter
${\boldsymbol w}_k[i]$. The expressions in (\ref{avect}) and
(\ref{wvect}) require matrix inversions with cubic complexity (
$O((J)^3)$ and $O((G(n_r+1))^3)$.

\subsection{Transmit Diversity Selection and Relay Selection}

{  In this subsection, we explore the idea of transmit diversity
selection (TDS) and relay selection (RS) and how they can be used to
improve the performance of cooperative systems. In cooperative
wireless systems with multiple relays, there are links that have
very poor propagation conditions that can degrade the performance of
the overall system. These links can be identified and removed from
the operation of the system via TDS and RS. To this end, we
formulate a TDS and RS strategy for a $2$ DF MIMO network as a
discrete combinatorial MSE problem which optimizes the use of the
channels of the second phase via $1$-bit power allocation
\cite{TDS_clarke}. It turns out that the problems of TDS and RS are
combinatorial problems which require either an exhaustive search or
some relaxation approach. We specify that a subset of $K_{sub}$
antennas of the $n_{r}K$ relay antennas are active at each time
instant in order to reduce the optimization complexity but also to
ensure a minimum available level of diversity. The destination
node's MSE TDS optimization function is given by }
\begin{equation}
\begin{split}
\boldsymbol{\mathcal{T}}_{r}^{opt} & =
\underset{\boldsymbol{\mathcal{T}}_{r}\in \Omega_{T}}{\arg\ \min}\
\mathcal{C}\big[i,\boldsymbol{\mathcal{T}}_{r},\boldsymbol{r}\big]\\
& = \underset{\boldsymbol{\mathcal{T}}_{r}\in \Omega_{T}}{\arg\
\min}\ \sum^{K}_{k=1} E \Big[ \big\Vert
b_{k}[i]-\boldsymbol{w}_{k}[i]
\boldsymbol{r}[i]\big\Vert^{2}\Big], \label{eq:mmse_opt_function}
\end{split}
\end{equation}
where
$\boldsymbol{\mathcal{T}}_{r}=\mathrm{diag}(a^{1}_{r_{1}d}\hdots
a^{K}_{r_{1}d},\hdots,a^{1}_{r_{n_{r}}d}\hdots a^{K}_{r_{n_{r}}d})$
and $a^{k}_{r_{j}d} = \{0,1\}$, and $\boldsymbol{w}_{k}$ is the
linear MMSE filter for the $k^{th}$ symbol. Under the assumption of
no inter-relay communication and that each data stream is allocated
to its correspondingly numbered transmit antenna at each relay, the
set $\Omega_{T}$ has a cardinality of
$|\Omega_{T}|={{n_{r}K}\choose{K_{sub}}}$ and contains all possible
combinations of relay transmit antennas patterns. The performance
and complexity of solutions to (\ref{eq:mmse_opt_function}) depend
on $|\Omega_{T}|$ and its elements. However, $|\Omega_{T}|$ is
significant even for modest numbers of antennas and relays, e.g.
$n_{r}\geq4$ and $K\geq2$. Further improvements can be achieved by a
process we term RS which addresses the possibility of mismatching
poor first phase channels with optimized second phase channels as
well as reducing the cardinality of $\Omega_{T}$.

By removing one or more relays based on their MSE
performance from consideration by (\ref{eq:mmse_opt_function}),
$\Omega_{T}$ can be optimized and its cardinality improved without
overly restricting the second-phase channels available to the TDS
process. The selection of the single highest MSE relay can be
expressed as a discrete maximization problem given by
\begin{multline}
j_{opt} = \underset{j \in \Omega_{R}}{\arg\ \max}\ \mathcal{F}\big[i,\boldsymbol{r}_{sr_{j}},\big]\\
=\underset{j \in \Omega_{\mathrm{R}}}{\arg\ \max} \sum^{K}_{k=1} E\Big[\big\Vert b_{k}[i]-\boldsymbol{w}_{j,k}[i] \boldsymbol{r}_{sr_{j}}[i]\big\Vert^{2}\Big],
\label{eq:discrete_opt_RS}
\end{multline}
where $\Omega_{\mathrm{R}}$ is the set of candidate relays and
$\boldsymbol{w}_{j,k}[i]$ is the MMSE filter for the $k^{th}$
symbol at the $j^{th}$ relay. On the solution of
(\ref{eq:discrete_opt_RS}), a refined subset, $\bar{\Omega}_{T}\in
\Omega_{T}$, is generated by removing members of $\Omega_{T}$
which involve transmission from relay $j_{opt}$, i.e. members of
$\Omega_{T}$ where $[a^{1}_{r_{j_{opt}}d}\hdots
a^{K}_{r_{j_{opt}}d}]\neq 0$. TDS then operates with this subset,
where $|\bar{\Omega}_{T}|={{K(n_{r}-1)}\choose{K_{sub}}}$.
Extension to the selection of multiple relays involves summing the
MSE from candidate relays and populating $\Omega_{R}$ with sets of
these relays. However, the selection of the number of relays to
remove is vital, as too high a value will result in a overly
restricting the second phase channels available to the TDS process
therefore increasing the probability of a channel mismatch.

\subsection{Cooperative MMSE Channel Estimation}

{  The next task that is necessary for the interference mitigation
and resource allocation is to compute the channel gains of the links
of the cooperative system. In order to estimate the channel in the
cooperative system under study, let us first consider the
transmitted signal for user $k$, ${\boldsymbol x}_{k}[i] =
\widetilde{\boldsymbol B}_k[i] \widetilde{\boldsymbol A}_k[i]
\widetilde{\boldsymbol {\mathcal C}}_k {\boldsymbol h}_k[i]=
{\boldsymbol Q}_k[i]{\boldsymbol h}_k[i]$, and the covariance matrix
given by}
\begin{equation}
\begin{split}
{\boldsymbol R}& =[{\boldsymbol r}[i] {\boldsymbol r}^H[i]] \\ & =
\sum_{k=1}^{K} {\boldsymbol Q}_k[i] E[ {\boldsymbol h}_k[i]
{\boldsymbol h}_k^H[i]]{\boldsymbol Q}_k^H[i] + E[{\boldsymbol
\eta}[i] {\boldsymbol \eta}^H[i]] + \sigma^2 {\boldsymbol I} \\ &
= \sum_{k=1}^{K} {\boldsymbol Q}_k[i] {\boldsymbol
P}_{{\boldsymbol h}_k} {\boldsymbol Q}_k^H[i] + {\boldsymbol
P}_{\eta} + \sigma^2 {\boldsymbol I}
\end{split}
\end{equation}
A linear estimator of ${\boldsymbol h}_k[i]$ applied to
${\boldsymbol r}[i]$ can be represented as $\hat{\boldsymbol
h}_k[i] = {\boldsymbol F}^H_k{\boldsymbol r}[i]$. The linear MMSE
channel estimation problem for the cooperative system under
consideration is formulated as
\begin{equation}
\begin{split}
{\boldsymbol F}_{k,{\rm opt}} & = \arg \min_{{\boldsymbol F}_k} E
\big[ ||{\boldsymbol h}_k[i] - \hat{\boldsymbol h}_k[i] ||^2 \big]
\\ & = \arg \min_{{\boldsymbol T}_k} E
\big[ ||{\boldsymbol h}_k[i] - {\boldsymbol T}^H_k{\boldsymbol
r}[i] ||^2 \big].
\end{split}
\end{equation}
Computing the gradient terms of the argument and equating them to
zero yields the MMSE solution
\begin{equation}
\begin{split}
{\boldsymbol F}_{k,{\rm opt}} & = {\boldsymbol R}^{-1}
{\boldsymbol P}_{k},
\end{split}
\end{equation}
where ${\boldsymbol P}_{k} = E[{\boldsymbol r}[i]{\boldsymbol
h}_k^H[i]] =  {\boldsymbol Q}_k[i] E[ {\boldsymbol h}_k[i]
{\boldsymbol h}_k^H[i]]=  {\boldsymbol Q}_k[i] {\boldsymbol
P}_{{\boldsymbol h}_k} $. Using the relation $\hat{\boldsymbol
h}_k[i] = {\boldsymbol F}^H_k{\boldsymbol r}[i]$, we obtain
\begin{equation}
\begin{split}
\hat{\boldsymbol h}_k[i] & = {\boldsymbol F}^H_{k,{\rm
opt}}{\boldsymbol r}[i] = {\boldsymbol P}_{k}^H{\boldsymbol
R}^{-1}{\boldsymbol r}[i]\\ & =  {\boldsymbol P}_{{\boldsymbol
h}_k}^H{\boldsymbol Q}_k^H [i]\big(\sum_{k=1}^{K} {\boldsymbol
Q}_k[i] {\boldsymbol P}_{{\boldsymbol h}_k}{\boldsymbol Q}_k^H[i]
+ {\boldsymbol P}_{\eta} + \sigma^2 {\boldsymbol I}
\big)^{-1}{\boldsymbol r}[i],
\end{split}
\label{cest}
\end{equation}
The expressions in (\ref{cest}) require matrix inversions with
cubic complexity ( $O(J^3)$), however, this matrix inversion is
common to (\ref{wvect}) and needs to be computed only once for
both expressions. In what follows, computationally efficient
algorithms with quadratic complexity ($O(J^2)$) based on an
alternating optimization strategy will be detailed.

\section{Adaptive Algorithms}

{  In this section, we present algorithms to compute the parameters
of interest and the expressions derived in the previous section with
lower computational complexity. Specifically, we develop adaptive
RALS algorithms using a method to build the group of $G$ users based
on the power levels, and then we employ an alternating optimization
strategy for efficiently estimating the parameters of the receive
filters, the power allocation vectors and the channels. Despite the
joint optimization that is associated with a non-convex problem, the
proposed RALS algorithms have been extensively tested and have not
presented problems with local minima.}

\subsection{Group Allocation and Channel Estimation}

{  The first step in the proposed strategy is to build the group of
$G$ users that will be used for the power allocation and receive
filter design. A RAKE receiver \cite{ziemer}, which is equivalent to
a filter matched to the signature sequence of the desired user or
the spatial signature of the desired data stream will be used for
the group allocation. The RAKE receiver is employed to obtain
$z_k^{\rm RAKE}[i] = (\widetilde{\boldsymbol C}_k\hat{\boldsymbol
h}_k[i])^H{\boldsymbol r}[i]=\hat{\boldsymbol p}_k^H[i]{\boldsymbol
r}[i]$. The group is then formed according to}
\begin{equation}
{\rm compute} ~~{\rm the}~~G~~{\rm largest}~~|z_k^{\rm
RAKE}[i]|,~~ k=1,2, \ldots, K. \label{group}
\end{equation}
The design of the RAKE and the other tasks require channel
estimation. The power allocation, receive filter design and
channel estimation expressions given in (\ref{avect}),
(\ref{wvect}) and (\ref{cest}), respectively, are solved by
replacing the expected values with time averages, and RLS-type
algorithms with an alternating optimization strategy. In order to
solve (\ref{cest}) efficiently, we develop a variant of the RLS
algorithm that is described by
\begin{equation}
\hat{\boldsymbol h}_k[i] = \hat{\boldsymbol P}_{{\boldsymbol
h}_k}^H[i] {\boldsymbol Q}_k^H[i] \hat{\boldsymbol R}^{-1}[i]
{\boldsymbol r}[i], \label{cestrec}
\end{equation}
where ${\boldsymbol Q}_k[i] = \widetilde{\boldsymbol B}_k[i]
\widetilde{\boldsymbol A}_k[i] \widetilde{\boldsymbol {\mathcal
C}}_k $, the estimate of the inverse of the covariance matrix
$\hat{\boldsymbol R}^{-1}[i]$ is computed with the matrix
inversion lemma \cite{haykin}
\begin{equation}
{\boldsymbol k}[i] = \frac{\alpha^{-1} \hat{\boldsymbol R}[i-1]
{\boldsymbol r}[i]}{1+\alpha^{-1} {\boldsymbol r}^H[i]
\hat{\boldsymbol R}[i-1] {\boldsymbol r}[i]}\label{kgain},
\end{equation}
\begin{equation}
\hat{\boldsymbol R}[i] = \alpha^{-1} \hat{\boldsymbol R}[i-1] -
\alpha^{-1} {\boldsymbol k}[i] {\boldsymbol r}^H[i]
\hat{\boldsymbol R}[i-1], \label{mil2}
\end{equation}
and
\begin{equation}
\hat{\boldsymbol P}_{{\boldsymbol h}_k}[i] = \alpha
\hat{\boldsymbol P}_{{\boldsymbol h}_k}[i-1] + \hat{\boldsymbol
h}_k[i-1] \hat{\boldsymbol h}_k^H[i-1] \label{Prec},
\end{equation}
where $\alpha$ is a forgetting factor that should be close to but
less than $1$.

\subsection{Joint Interference Suppression and Power Allocation}

The approach for allocating the power within a
group is to drop the constraint, estimate the quantities of
interest and then impose the constraint via a subsequent
normalization. The group-based power allocation algorithm
is computed by
\begin{equation}
\begin{split}
\hat{\boldsymbol a}_{{\mathbf{\mathcal S}},k}[i] & =
\hat{\boldsymbol R}_{{\mathbf{\mathcal S}},k}[i] \hat{\boldsymbol
p}_{{\mathbf{\mathcal S}},k}[i]\\ & = \hat{\boldsymbol
R}_{{\mathbf{\mathcal S}},k}[i] (\alpha\hat{\boldsymbol
p}_{{\mathbf{\mathcal S}},k}[i-1] + b_k[i] {\boldsymbol
v}_k[i])\\& =  \hat{\boldsymbol a}_{{\mathbf{\mathcal S}},k}[i-1]
+ \xi_{a}[i] {\boldsymbol k}_{{\mathbf{\mathcal S}},k}[i],
\label{arls}
\end{split}
\end{equation}
where $\xi_a[i] = b_k[i] - \hat{\boldsymbol a}_{{\mathbf{\mathcal
S}},k}^H[i-1] {\boldsymbol v}_k[i]$ is the a priori error,
${\boldsymbol v}_k[i] = {\boldsymbol B}_{\mathbf {\mathcal
S}}^H[i]{\boldsymbol P}_{\mathbf {\mathcal S}}^H[i] {\boldsymbol
w}_k[i]$ is the input signal to the recursion
\begin{equation}
{\boldsymbol k}_{{\mathbf{\mathcal S}},k}[i] = \frac{\alpha^{-1}
\hat{\boldsymbol R}_{{\mathbf{\mathcal S}},k}[i-1] {\boldsymbol
v}_k[i]}{1+\alpha^{-1} {\boldsymbol v}_k^H[i] \hat{\boldsymbol
R}_{{\mathbf{\mathcal S}},k}[i-1] {\boldsymbol v}_k[i] },
\end{equation}
\begin{equation}
\hat{\boldsymbol R}_{{\mathbf{\mathcal S}},k}[i] = \alpha^{-1}
\hat{\boldsymbol R}_{{\mathbf{\mathcal S}},k}[i-1] - \alpha^{-1}
 {\boldsymbol k}_{{\mathbf{\mathcal S}},k}[i] {\boldsymbol
v}_k^H[i] \hat{\boldsymbol R}_{{\mathbf{\mathcal S}},k}[i-1].
\end{equation}
The normalization $\hat{\boldsymbol a}_{{\mathbf{\mathcal
S}},k}[i] \leftarrow P_G  ~\hat{\boldsymbol a}_{{\mathbf{\mathcal
S}},k}[i]/||\hat{\boldsymbol a}_{{\mathbf{\mathcal S}},k}[i]|| $
is then performed to ensure the power constraint.

The linear receive filter is computed by
\begin{equation}
\hat{\boldsymbol w}_k[i] = \hat{\boldsymbol w}_k[i-1] +
{\boldsymbol k}[i] \xi^*[i], \label{wrls}
\end{equation}
where the a priori error is given by $\xi[i] = b_k[i] -
\hat{\boldsymbol w}_k^H[i-1] {\boldsymbol r}[i]$ and ${\boldsymbol
k}[i]$ is given by (\ref{kgain}). The proposed scheme employs the
algorithm in (\ref{group}) to allocate the users in the group and
the channel estimation approach of (\ref{cestrec})-(\ref{Prec}).
The alternating optimization strategy uses the recursions
(\ref{arls}) and (\ref{wrls}) with $1~{\rm or}~2$ iterations per
symbol $i$.

\subsection{Transmit Diversity Selection and Relay Selection Based on Discrete Stochastic Gradient Algorithms}

{  In this part, we describe a low-complexity solution to the joint
TDS and RS problem based on a discrete stochastic gradient algorithm
(DSA) that can compute the optimal combinatorial solution outline in
(\ref{eq:mmse_opt_function}) and (\ref{eq:discrete_opt_RS}) with a
substantially reduced cost as compared with the exhaustive search.
We present a pair of low-complexity DSA that was first reported in
\cite{global_search_discrete_opt_Andradottir,MIMO_antenna_discrete_optimization_Krishnamurthy},
which jointly optimizes RS and TDS in accordance with
(\ref{eq:mmse_opt_function}) and (\ref{eq:discrete_opt_RS}), and
converges to the optimal exhaustive solution.}

\begin{table}[h]
\caption{Proposed discrete stochastic joint TDS and RS algorithm}
\begin{spacing}{1}
\begin{footnotesize}
\begin{tabular}{l}
\toprule
\textbf{Step}\\
\textbf{1. Initialization}\\
\hspace{0.3cm} choose $j[1] \in \Omega_{\mathrm{R}}, j^{W}[1] \in\Omega_{R}$,
$\boldsymbol{\pi}_{R}\big[1,j[1]\big]=1$,
$\boldsymbol{\pi}_{R}[1,\bar{j}]=0$ \\
\hspace{0.3cm}for $\bar{j}\neq j[1]$\\
\textbf{2. For the time index} $i=1,2, ... , N$\\
\hspace{0.3cm} choose $j^{C}[i] \in \Omega_{R}$\\
\textbf{3. Comparison and update of the worst performing relay}\\
\hspace{0.3cm}if {$\mathcal{F}\big[i,\boldsymbol{r}_{sr_{j^{C}[i]}}\big] > \mathcal{F}\big[i,\boldsymbol{r}_{sr_{j^{W}[i]}}\big]$}
then $j^{W}[i+1] = j^{C}[i]$\\
\hspace{0.3cm}otherwise $j^{W}[i+1] = j^{W}[i]$\\
\textbf{4. State occupation probability (SOP) vector update}\\
\hspace{0.3cm}$\boldsymbol{\pi}_{R}[i+1] = \boldsymbol{\pi}_{R}[i] + \mu[i+1](\mathbf{v}_{j^{W}[i+1]}-\boldsymbol{\pi}_{R}[i])$ where $\mu[i] = 1/i$\\
\textbf{5. Determine largest SOP vector element and select the optimum relay}\\
\hspace{0.3cm}if {$\boldsymbol{\pi}_{R}\big[i+1,j^{W}[i+1]\big] > \boldsymbol{\pi}_{R}[i+1,j[i]]$}
then $j[i+1] = j^{W}[i+1]$\\
\hspace{0.3cm} otherwise $j[i+1] = j[i]$\\
\textbf{6. TDS Set Reduction}\\
\hspace{0.3cm} remove members of $\Omega_{T}$  which utilize relay $j[i+1]$ ($\Omega_{T} \rightarrow \bar{\Omega}_{T})$\\
\bottomrule
\end{tabular}
\end{footnotesize}
\end{spacing}
\label{tab:RS_algorithm}
\end{table}

The RS portion of the DSA is given by the algorithm of Table
\ref{tab:RS_algorithm}. At each iteration the MSE of a randomly
chosen candidate relay ($j^{C}$) (step 2) and that of the worst
performing relay currently known ($j^{W}$) are calculated (step 3).
Via a comparison, the higher MSE relay is designated $j^{W}$ for the
next iteration (step 3). The current solution and the relay chosen
for removal ($j$) is denoted as the current optimum and is the relay
which has occupied $j^{W}$ most frequently over the course of the
packet up to the $i^{th}$ time instant; effectively an average of
the occupiers of $j^{W}$. This averaging/selection process is
performed by allocating each member of $\Omega_{R}$ a $|\Omega_{R}|
\times 1$ unit vector, $\mathbf{v}_{l}$, which has a one in its
corresponding position in $\Omega_{R}$, i.e.,
$\mathbf{v}_{j^{W}}[i]$ is the label of the worst performing relay
at the $i^{th}$ iteration. The current optimum is then chosen and
tracked by means of a $| \Omega_{R}| \times 1$ state occupation
probability (SOP) vector, $\boldsymbol{\pi}_{R}$. This vector is
updated at each iteration by adding $\mathbf{v}_{j^{W}}[i+1]$ and
subtracting the previous value of $\boldsymbol{\pi}_{R}$ (step 4).
The current optimum is then determined by selecting the largest
element in $\boldsymbol{\pi}_{R}$ and its corresponding entry in
$\Omega_{R}$ (step 5). Through this process, the current optimum
converges towards and tracks the exhaustive solution
\cite{global_search_discrete_opt_Andradottir}. An alternative
interpretation of the proposed algorithm is to view the transitions,
$j^{W}[i]\rightarrow j^{W}[i+1]$, as a Markov chain and the members
of $\Omega_{R}$ as the possible transition states. The current
optimum can then be defined as the most visited state.

Once RS is complete at each time instant, set reduction
($\Omega_{T}\rightarrow\bar{\Omega}_{T}$, step 6) and TDS can take
place. To perform TDS, modified versions of steps $1-5$ are used.
The considered set is replaced, $\Omega_{R} \rightarrow
\bar{\Omega}_{T}$; the structure of interest is replaced, $j
\rightarrow \boldsymbol{\mathcal{T}}_{r}$; the best performing
matrix is sought $j^{W} \rightarrow
\boldsymbol{\mathcal{T}}_{r}^{B}$; the SOP vector is replaced
$\boldsymbol{\pi}_{R}\rightarrow
\boldsymbol{\pi}_{\boldsymbol{\mathcal{T}}}$ and
$\mathcal{C}\rightarrow \mathcal{F}$ from
(\ref{eq:mmse_opt_function}). Finally, the inequality of step 3 is
reversed to enable convergence to the lowest MSE TDS matrix which is
then feedback to the relays in the form of 1-bit per relay antenna.



Significant complexity savings result from the proposed algorithm; savings which increase with $K$, $n_{r}$ and the number relays removed in the RS process. For example, when $n_{r} = 10$, $K=2$ and 4 relays are removed, the number of complex multiplications for MMSE reception and exhaustive TDS, exhaustive TDS with RS, iterative TDS and iterative TDS with RS are $5.8\times 10^{8}$, $1.7\times 10^{8}$, $1.8\times 10^{5}$ and $5.9\times 10^{4}$, respectively, for each time instant.

\section{Analysis and Requirements of the Algorithms}

{  In this section, we assess the requirements of the proposed and
existing algorithms for cross-layer design in terms of computational
complexity and number of feedback bits. The basic idea is to show
the computational cost of the algorithms presented and compare them
with those of existing techniques for interference mitigation and/or
resource allocation.}

\subsection{Computational Complexity Requirements}

We discuss here the computational complexity of the proposed and
existing algorithms. Specifically, we will detail the required
number of complex additions and multiplications of the proposed
JPAIS-GBC algorithms and compare them with
interference suppression schemes without cooperation (NCIS) and
with cooperation (CIS) using an equal power allocation across the
relays. Both uplink and downlink scenarios are considered in the
analysis. In Table I we show the computational complexity required
by each recursion associated with a parameter vector/matrix for
the JPAIS-GBC with $G=K$, which is more suitable for the uplink.

\begin{table}[h]
\centering%
\caption{\small Computational complexity of algorithms with a
global power constraint $G=K$ .} {
\begin{tabular}{ccc}
\hline \rule{0cm}{2.5ex}&  \multicolumn{2}{c}{\small Number of
operations per symbol } \\ \cline{2-3} {\small Parameter} & {\small
Additions} & {\small Multiplications} \\ \hline
\emph{ \bf  } & {\small $2(J)^{2} $} & {\small $3(J)^{2}$}   \\
\emph{\bf \small $\hat{\boldsymbol W}[i]$} & {\small $ + 2K(J)$} & {\small $ + 2K(J)$}   \\
\emph{\bf } & {\small $ - J+1$} & {\small $+ 3J+1$}
\\ \hline
\emph{\small \bf }  & {\small $3K(K(n_r+1))$} & {\small $K(K(n_r+1))^2$} \\
\emph{}  & {\small $+K(n_r+1)(L-1)$} & {\small $+4(K(n_r+1))^2$}
\\
\emph{\small \bf $\hat{\boldsymbol a}_T[i]$ }  & {\small $+K(M(n_r+1))$} & {\small $+(K+L)(K(n_r+1))^2$} \\
\emph{ }  & {\small $+K(K(n_r+1))$} & {\small $-(K(n_r+1))^2$}\\
\emph{}  & {\small $+6(K(n_r+1))^2 $} & {\small $+K(MQ)$} \\
\emph{}  & {\small $ + 3K(n_r+1)+n_r+2$} & {\small $+n_r$}
\\\hline
\emph{}  & {\small $5(KQ)^{2} $} & {\small $+5(K(n_r+1)^{2} $} \\
\emph{\small \bf $\hat{\boldsymbol h}_k[i]$ }  & {\small $+5KQ$} & {\small $+6KQ$} \\
\emph{}  & {\small $+ 3$} & {\small $+1$} \\  \hline
\end{tabular}
}
\end{table}

In Table II we show the computational complexity required by each
recursion associated with a parameter vector for the JPAIS-GBC
algorithm, which is suitable for both the uplink and the downlink.
A noticeable difference between the JPAIS-GBC with $G=K$ and $G=1$ is
that the latter is employed for each user, whereas the former is
used for all the $K$ users in the system. Since the computation of
the inverse of $\hat{\boldsymbol R}[i]$ is common to all users for
the uplink in our system, the JPAIS-GBC with $G=K$ is more efficient than the
JPAIS-GBC with $G=1$ computed for all the $K$ users.

\begin{table}[h]
\centering%
\caption{\small Computational complexity of algorithms with
individual power constraints ($G=1$).} {
\begin{tabular}{ccc}
\hline \rule{0cm}{1.5ex}&  \multicolumn{2}{c}{\small Number of
operations per symbol } \\  \cline{2-3} {\small Parameter} & {\small
Additions} & {\small Multiplications} \\ \hline
\emph{ \bf  } & {\small $2(J)^{2} $} & {\small $3(J)^{2}$}   \\
\emph{\bf \small $\hat{\boldsymbol w}_k[i]$} & {\small $ + J$} & {\small $ + 5J$}   \\
\emph{\bf } & {\small $ +1$} & {\small $+1$}
\\ \hline
\emph{\small \bf }  & {\small $2(n_r+1)^2$} & {\small $3(n_r+1)^2$} \\
\emph{}  & {\small $+3(n_r+1)$} & {\small $+7(n_r+1)$} \\
\emph{\small \bf $\hat{\boldsymbol a}_k[i]$ }  & {\small $+JL$} & {\small $+JL$} \\
\emph{ }  & {\small $+Q$} & {\small $+Q$}\\
\emph{}  & {\small $-3 $} & {\small $+3$}
\\\hline
\emph{}  & {\small $2(Q)^{2} $} & {\small $6(Q)^{2} $} \\
\emph{\small \bf $\hat{\boldsymbol h}_k[i]$ }  & {\small $+5MQ$} & {\small $+MQ$} \\
\emph{}  & {\small $-5(n_r+1)+ 3$} & {\small $+4(n_r+1)+1$} \\
\hline
\end{tabular}
}
\end{table}

The recursions employed for the proposed JPAIS-GBC with $G=K$ and the
JPAIS-GBC with $G=1$ are general and parts of them are used in the existing
CIS and NIS algorithms. Therefore, we can use them to describe the
required computational complexity of the existing algorithms. In
Table III we show the required recursions for the proposed and
existing algorithms, whose complexity is detailed in Tables I and
II.

\begin{table}[h]
\centering%
\caption{\small Computational complexity of the proposed JPAIS and
existing algorithms.} {
\begin{tabular}{ll}
\hline \rule{0cm}{2.5ex} {\small Algorithm} & {\small Recursions }
\\ \hline
\emph{\small \bf  JPAIS-GPC($G=K$) ~(Uplink) } & {\small $\hat{\boldsymbol W}[i]$, $\hat{\boldsymbol a}_T[i]$,$\hat{\boldsymbol h}_k[i]$}   \\
 \hline
\emph{\small \bf JPAIS-GBC ($G=1$)~(Downlink) }  & {\small $\hat{\boldsymbol w}_k[i]$, $\hat{\boldsymbol a}_k[i]$, $\hat{\boldsymbol h}_k[i]$ }  \\
\hline
\emph{\small \bf CIS (Uplink) }  & {\small $\hat{\boldsymbol W}[i]$, $\hat{\boldsymbol a}_T[i]$ is fixed}  \\
\hline
\emph{\small \bf CIS (Downlink) }  & {\small $\hat{\boldsymbol w}_k[i]$, $\hat{\boldsymbol a}_k[i]$ is fixed}  \\
 \hline
\emph{\small \bf NCIS (Uplink)}  & {\small $\hat{\boldsymbol W}[i]$ with $n_r=0$}  \\
 \hline
\emph{\small \bf NCIS (Downlink)}  & {\small $\hat{\boldsymbol w}_k[i]$ with $n_r=0$}  \\
\hline

\end{tabular}
}
\end{table}

\begin{figure}[!htb]
\begin{center}
\def\epsfsize#1#2{1.00\columnwidth}
\epsfbox{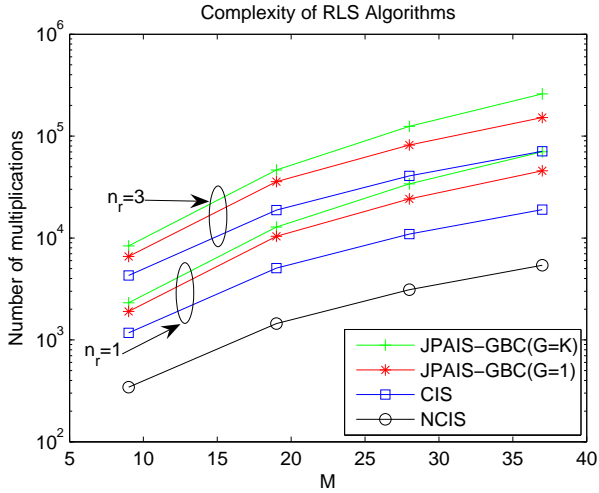} \vspace{-1.0em}\caption{\footnotesize
Computational complexity in terms of the number of complex
multiplications of the proposed and existing schemes for the
uplink.} \vspace{-0.25em}\label{fig_comp}
\end{center}
\end{figure}

In Fig. \ref{fig_comp}, we illustrate the required computational
complexity for the proposed and existing schemes for different
number of relays ($n_r$). The curves show that the proposed
JPAIS-GBC with $G=K$ and JPAIS-GBC with $G=1$ are more complex than the CIS scheme and
the NCIS.  This is due to the fact that the power allocation and
channel estimation recursions are employed. However, we will show
in the next section that this additional required complexity
(which is modest) can significantly improve the performance of the
system.

\subsection{Feedback Channel Requirements}

\begin{figure}[!htb]
\begin{center}
\def\epsfsize#1#2{1\columnwidth}
\epsfbox{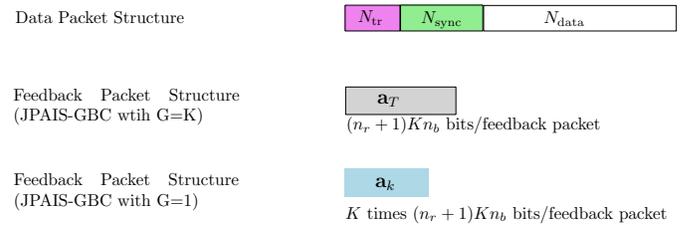} \vspace{-0.25em} \caption{\footnotesize Proposed
structure of the data and feedback packets.} \label{fig_packet}
\end{center}
\end{figure}

{  The JPAIS algorithms presented so far for cross-layer design
require feedback signalling in order to allocate the power levels
across the relays.} In order to illustrate how these requirements
are addressed, we can refer to Fig. \ref{fig_packet} which depicts
the structure for both the data and feedback packets. The data
packet comprises a number of allocated bits for training ($N_{\rm
tr}$), for synchronization and control ($N_{\rm sync}$) and the
transmitted data ($N_{\rm data}$). The feedback packet requires the
transmission of the power allocation vector ${\boldsymbol a}_T$ for
the case of the JPAIS-GBC algorithm with $G=K$, whereas it requires
the transmission of ${\boldsymbol a}_k$ for each user for JPAIS-GBC
with $G=1$. A typical number of bits $n_b$ required to quantize each
coefficient of the vectors ${\boldsymbol a}_T$ and ${\boldsymbol
a}_k$ via scalar quantization is $n_b=4$ bits. More efficient
schemes employing vector quantization \cite{gersho,delamare_ieeproc}
and that take into account correlations between the coefficients are
also possible.

For the uplink (or multiple-access channel), the base station (or
access point) needs to feedback the power levels across the links
to the $K$ destination users in the system. With the JPAIS-GBC with $G=K$
algorithm, the parameter vector ${\boldsymbol a}_T$ with $(n_r+1)K
n_b$ bits/packet must be broadcasted to the $K$ users. For the
JPAIS-GBC algorithm with $G=1$, a parameter vector ${\boldsymbol a}_k$ with
$(n_r+1) n_b$ bits/packet must be broadcasted to each user in the
systems. In terms of feedback, the JPAIS-GBC algorithm with $G=1$ is more
flexible and may require less feedback bits if there is no need
for a constant update of the power levels for all $K$ users.

For the downlink (or broadcast channel), the $K$ users must
feedback the power levels across the links to the base station.
With the JPAIS-GBC algorithm with $G=K$, the parameter vector ${\boldsymbol
a}_T$ with $(n_r+1)K n_b$ bits/packet must be computed by each
user and transmitted to the base station, which uses the
${\boldsymbol a}_T$ vector coming from each user. An algorithm for
data fusion or a simple averaging procedure can be used. For the
JPAIS-GBC algorithm with $G=1$, a parameter vector ${\boldsymbol a}_k$ with
$(n_r+1) n_b$ bits/packet must be transmitted from each user to
the base station. In terms of feedback, the JPAIS-GBC algorithm with $G=1$
requires significantly less feedback bits than the JPAIS-GBC with $G=K$ in
this scenario.

The MIMO TDS and RS scheme can be interpreted as a $1$-bit power
allocation scheme and therefore achieves performance improvements
whilst utilizing the minimum number of feedback bits per antenna.
Consequently, the feedback requirements per update of a cooperative
MIMO system using TDS and RS is given by the total number of relay
transmit antennas $n_{r}K$. This minimal feedback allows optimization
of the system whilst maintaining the capacity of the system with
regards to the transmission of useful data.

\section{Simulations}

{  In this section, we illustrate with Monte-Carlo simulations the
performance of the cross-layer algorithms described in this chapter.
Specifically, we assess the performance in terms of the bit error
ratio (BER) of the JPAIS scheme and adaptive algorithms with
group-based power constraints (GBC). The JPAIS scheme and algorithms
are compared with schemes without cooperation (NCIS) and with
cooperation (CIS) \cite{venturino} using an equal power allocation
across the relays (the power allocation in the JPAIS scheme is
disabled).} We also assess the proposed algorithms for transmit
diversity selection and relay selection (Iterative TDS with RS) are
presented and comparisons drawn against the optimal exhaustive
solutions (Exhaustive TDS with RS), the unmodified system (No TDS),
and the direct transmission (Non-Cooperative).

\subsection{DS-CDMA System}

A DS-CDMA network with randomly generated spreading codes and a
processing gain $N=16$ is considered. The fading channels are
generated considering a random power delay profile with gains taken
from a complex Gaussian variable with unit variance and mean zero,
$L=5$ paths spaced by one chip, and are normalized for unit power.
The power constraint parameter $P_{A,k}$ is set for each user so
that the designer can control the SNR (${\rm SNR} =
P_{A,k}/\sigma^2$) and $P_T= P_G + (K-G) P_{A,k}$, whereas it
follows a log-normal distribution for the users with associated
standard deviation equal to $3$ dB. The DF cooperative protocol is
adopted and all the relays and the destination terminal use either
linear MMSE, which have full channel and noise variance knowledge,
or adaptive receivers. The receivers are adjusted with the proposed
RALS with $2$ iterations for the JPAIS scheme, and with RLS
algorithms for the NCIS and CIS schemes. We employ packets with
$1500$ QPSK symbols and average the curves over $1000$ runs. For the
adaptive receivers, we provide training sequences with $N_{\rm
tr}=200$ symbols placed at the preamble of the packets. After the
training sequence, the adaptive receivers are switched to
decision-directed mode.

\begin{figure}[!htb]
\begin{center}
\def\epsfsize#1#2{1\columnwidth}
\epsfbox{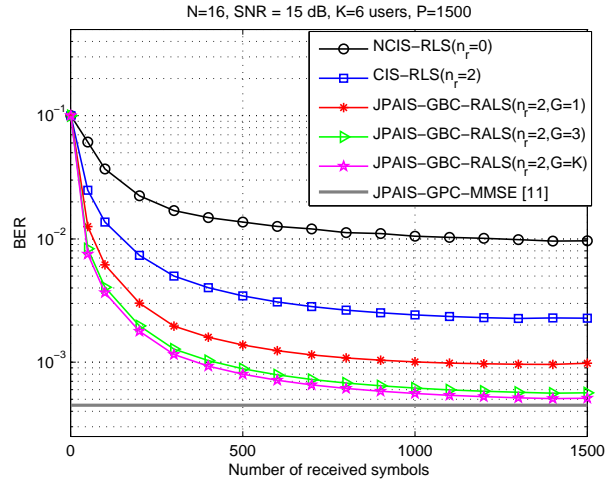} \vspace{-1.5em} \caption{\footnotesize BER
performance versus number of symbols. Parameters: AF protocol,
$\lambda_T=\lambda_k =0.025$ (for MMSE schemes), $\alpha=0.998$,
$\hat{\boldsymbol R}_{{\mathbf{\mathcal S}},k}^{-1}[i]=0.01
{\boldsymbol I}$ and $\hat{\boldsymbol R}^{-1}[i]=0.01 {\boldsymbol
I}$.}  \label{fig1}
\end{center}
\end{figure}

The first experiment depicted in Fig. \ref{fig1} shows the BER
performance of the proposed JPAIS scheme and algorithms against
the NCIS and CIS schemes with $n_r=2$ relays. The JPAIS scheme is
considered with the group-based power constraints (JPAIS-GBC). All
techniques employ MMSE or RLS-type algorithms for estimation of
the channels, the receive filters and the power allocation for
each user. The results show that as the group size $G$ is
increased the proposed JPAIS scheme and algorithms converge to
approximately the same level of the cooperative JPAIS-MMSE scheme
reported in \cite{delamare_jpais}, which employs $G=K$ for power
allocation, and has full knowledge of the channel and the noise
variance.

\begin{figure}[!htb]
\begin{center}
\def\epsfsize#1#2{1\columnwidth}
\epsfbox{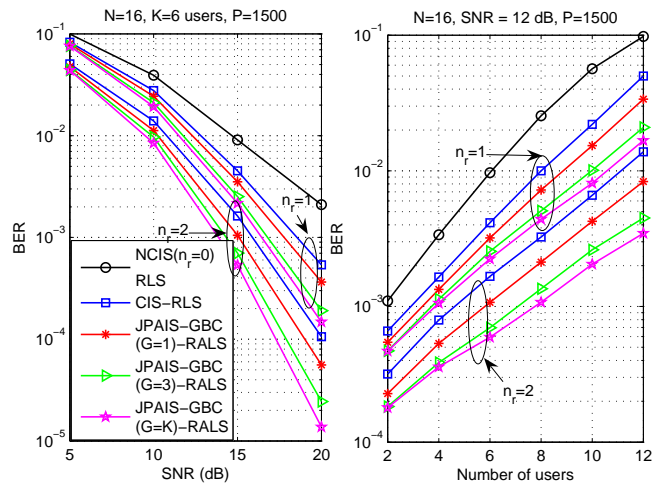} \vspace{-1.5em}\caption{\footnotesize BER
performance versus SNR and number of users for the optimal linear
MMSE detectors. Parameters: AF protocol, $\alpha =0.998$,
$\hat{\boldsymbol R}_{{\mathbf{\mathcal S}},k}^{-1}[i]=0.01
{\boldsymbol I}$ and $\hat{\boldsymbol R}^{-1}[i]=0.01 {\boldsymbol
I}$.}  \label{fig2}
\end{center}
\end{figure}

The proposed JPAIS-GBC scheme is then compared with a
non-cooperative approach (NCIS) and a cooperative scheme with
equal power allocation (CIS) across the relays for $n_r=1,2$
relays. The results shown in Fig. \ref{fig2} illustrate the
performance improvement achieved by the JPAIS scheme and
algorithms, which significantly outperform the CIS and the NCIS
techniques. As the number of relays is increased so is the
performance, reflecting the exploitation of the spatial diversity.
In the scenario studied, the proposed JPAIS-GBC with $G=3$ can
accommodate up to $3$ more users as compared to the CIS scheme and
double the capacity as compared with the NCIS for the same BER
performance. The curves indicate that the GBC for power allocation
with only a few users is able to attain a performance close to the
JPAIS-GBC with $G=K$ users, while requiring a lower complexity and
less network signalling. A comprehensive study of the signalling
requirements will be considered in a future work.

\begin{figure}[!htb]
\begin{center}
\def\epsfsize#1#2{1\columnwidth}
\epsfbox{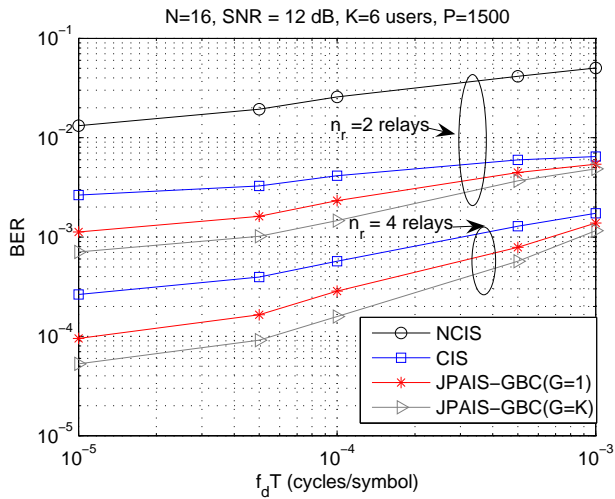} \vspace{-1.0em}\caption{\footnotesize BER
performance versus $f_dT$ for the AF protocol. The parameters of the
adaptive algorithms are optimized for each $f_dT$.}
\vspace{-0.15em}\label{fig6}
\end{center}
\end{figure}

The next experiment considers the average BER performance against
the normalized fading rate $f_dT$ (cycles/symbol), as depicted in
Fig. \ref{fig6}. The idea is to illustrate a situation where the
channel changes within a packet and the system transmits the power
allocation vectors computed by the proposed JPAIS algorithms via a
feedback channel. In this scenario, the JPAIS algorithms compute
the parameters of the receiver and the power allocation vector,
which is transmitted only once to the mobile users. This leads to
a situation in which the power allocation becomes outdated. The
results show that the gains of the proposed JPAIS algorithms
decrease gradually as the $f_dT$ is increased to the BER level of
the existing CIS algorithms for both $n_r=2$ and $n_r=4$ relays,
indicating that the power allocation is no longer able to provide
performance advantages. This problem requires the deployment of a
frequent update of the power allocation via feedback channels.

\begin{figure}[!htb]
\begin{center}
\def\epsfsize#1#2{1\columnwidth}
\epsfbox{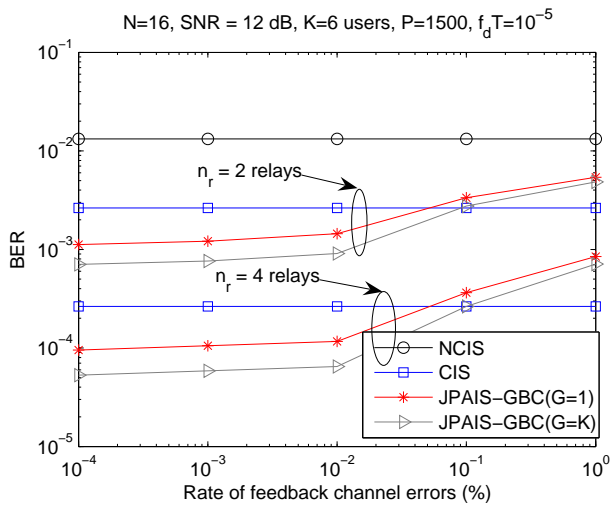} \vspace{-1.0em}\caption{\footnotesize BER
performance versus percentage of feedback errors.}
\vspace{-0.25em}\label{fig7}
\end{center}
\end{figure}

The last experiment, shown in Fig. \ref{fig7}, illustrates the
averaged BER performance versus the percentage of errors in the
feedback channel for an uplink scenario. Specifically, the
feedback packet structure is employed and each coefficient is
quantized with $4$ bits. Each feedback packet is constructed with
a sequence of binary data ($0$s and $1$s) and is transmitted over
a binary symmetric channel (BSC) with an associated probability of
error $P_e$. We then evaluate the BER of the proposed JPAIS and
the existing algorithms against several values of the $P_e$. The
results show that the proposed JPAIS algorithms obtain significant
gains over the existing CIS algorithm for values of $P_e < 0.1
\%$. As we increase the rate of feedback errors, the performance
of the proposed JPAIS becomes worse than the CIS algorithms. This
suggests the use of error-control coding techniques to keep the
level of errors in the feedback channel below a certain value.

\subsection{MIMO System}

In this part, simulations of the proposed algorithms (Iterative TDS with RS) are presented and comparisons drawn against the optimal exhaustive solutions (Exhaustive TDS with RS), the unmodified system where all antennas are active (No TDS), and the direct transmission (Non-Cooperative). Plots of the schemes with TDS only (Exhaustive TDS, Iterative TDS) are also included to illustrate the performance improvement obtained by RS. Equal power allocation is maintained in each phase so that the total transmit bit power of the relays is unity. RLS channel estimation (CE) is used where all auxiliary matrices are initialized as identity matrices and estimation matrices are zero matrices, and the exponential forgetting factor is 0.9. Each simulation is averaged over 1000 packets ($N_{\mathrm{p}}$), each with training sequences of $200$ symbols.

\begin{figure}[!htb]
\begin{center}
\def\epsfsize#1#2{1\columnwidth}
\epsfbox{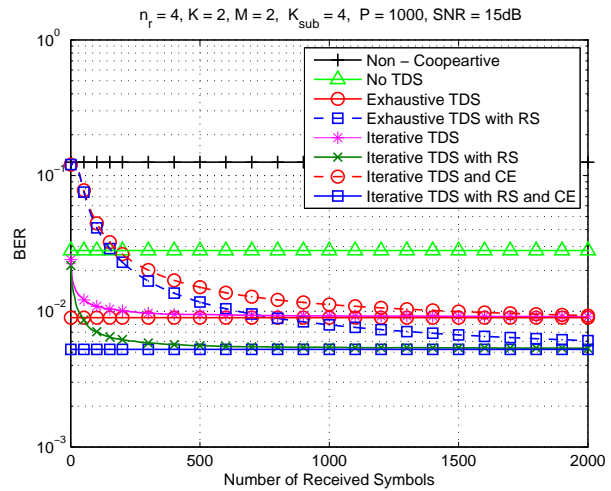} \caption{Cooperative DF MIMO BER performance
versus the number of received symbols.} \label{fig:ber_ce}
\end{center}
\end{figure}

Fig. \ref{fig:ber_ce} gives the BER convergence performance of the proposed algorithms. The iterative TDS with RS algorithm converges to the optimal BER as does TDS with RS and CE, albeit in a delayed fashion due to the CE. The TDS with RS scheme exhibits quicker convergence and lower steady state BER. These results and the interdependence between elements of the algorithm confirm that both the RS and TDS portions of the algorithm converge to their exhaustive solutions.

\begin{figure}[!htb]
\begin{center}
\def\epsfsize#1#2{1\columnwidth}
\epsfbox{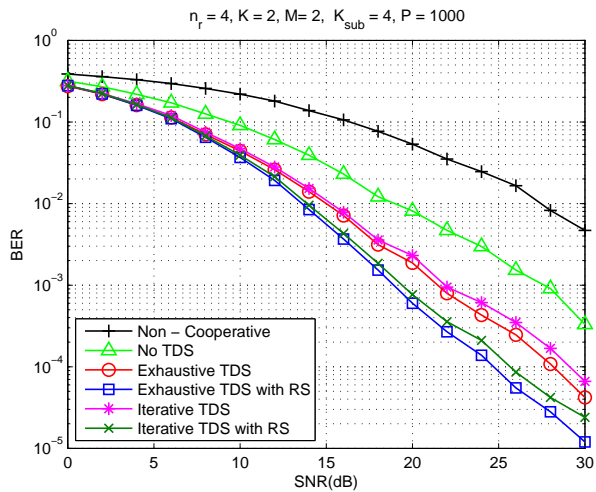} \caption{Cooperative FD MIMO BER performance
versus SNR.} \label{fig:SNR}
\end{center}
\end{figure}

Fig. \ref{fig:SNR} shows the BER versus SNR performance of the proposed and conventional algorithms. Increased diversity has been achieved without sacrificing multiplexing gain and illustrates that although the maximum available diversity advantage decreases from $M (n_{r}+1)$ to $M (K_{sub}/K+1)$ with RS with TDS because fewer antennas are active,  the actual diversity achieved has increased. These diversity effects can be attributed to the removal of poor paths and therefore a lower probability of first phase and second phase channel mismatch but also the increase in transmit power over the remaining paths. The largest gains in diversity are present in the
$15-25$dB region and begin to diminish above this region because relay decoding becomes increasingly reliable and lower power paths become more viable for transmission.

\section{Extensions and Suggestions for Future Work}

The algorithms for joint resource allocation and interference
mitigation described in this chapter are quite general and can be
employed in a variety of wireless communication systems that are
equipped with cooperative techniques. These include
orthogonal-frequency-division-multiplexing (OFDM) \cite{stuber},
single-carrier systems with frequency-domain equalisation (SC-FDE)
\cite{falconer} and ultra-wide band (UWB) systems \cite{moewin}.

Possible extensions include the incorporation of more advanced
interference mitigation strategies than linear schemes. These
include nonlinear detection techniques such as successive
interference cancellation \cite{vblast,delamare_itic}, decision
feedback strategies \cite{delamaretc,delamare_mber} and sphere
decoders \cite{hassibi}. The detection algorithms could also be
considered with space-time coding schemes \cite{alamouti,tarokh},
channel coding and iterative processing approaches
\cite{wang,delamaretc}.

Another complementary set of techniques comprises algorithms for
adaptive parameter estimation. These methods are fundamental to
estimate key parameters such as channel gains, amplitudes and
receive filters, whilst keeping the complexity low and being able to
track variations of the environment. Amongst the adaptive parameter
estimation techniques, a designer can choose between supervised and
blind approaches \cite{haykin}. Blind techniques
\cite{xutsa}-\cite{mswfccm} are appealing as they can increase the
spectral efficiency of wireless systems. This is especially relevant
for cooperative systems as they require extra signalling for
cross-layer design. Supervised adaptive algorithms usually rely on
training sequences that are sent at the beginning of each data
packet \cite{choi,hassibi2}. One fundamental issue in the choice of
the adaptive parameter estimation algorithm is the speed of
convergence and the tracking performance. The literature suggests
that reduced-rank algorithms \cite{delamaresp}-\cite{jidf},
\cite{goldstein,sun,avf} are very attractive choices when fast
training and accurate tracking are important issues.

\section{Concluding Remarks}

We have presented in this work joint iterative power allocation
and interference mitigation techniques for DS-CDMA and MIMO networks which
employ multiple hops and the AF and DF cooperation protocols. A joint
constrained optimization framework and algorithms that consider
the allocation of power levels across the relays subject to group
power constraints and the design of linear
receivers for interference suppression were proposed. A scheme
for joint transmit diversity optimisation and relay selection along
with linear interference suppression has also been detailed and
applied to MIMO systems. A study of
the requirements of the proposed and existing algorithms in terms
of computational complexity and feedback channels has also been
conducted. The results of simulations have shown that the proposed
algorithms obtain significant gains in performance and capacity
over existing non-cooperative and cooperative schemes.

\end{document}